\documentclass[11pt,a4paper]{article}
\usepackage{amsmath}
\usepackage[mathcal]{eucal}
\usepackage{latexsym}
\usepackage{amssymb}
\begin{titlepage} 
\title{Instanton representation of Plebanski gravity XXII. Minisuperspace quantization of gravity coupled to spin 1/2 fermionic fields}
\author{Eyo Eyo Ita III}
\medskip
\input amssym.def
\input amssym.tex

\def \in{\indent}

\begin{document}
\maketitle
\bigskip
\centerline{Department of Applied Mathematics and Theoretical Physics} 
\smallskip
\centerline{Centre for Mathematical Sciences, University of Cambridge, Wilberforce Road}
\smallskip
\centerline{Cambridge CB3 0WA, United Kingdom}
\smallskip
\centerline{eei20@cam.ac.uk} 

\bigskip

\begin{abstract}
In this paper we carry out a quantization of gravity coupled to massless spin 1/2 fermions through the instanton representation of Plebanski gravity.  We have constructed a Hilbert space of states for this model, and we have computed the Hamilton's equations of motion.  The classical equations appear, at a superficial level, to be consistent with the quantum dynamics of the theory.
\end{abstract}
\end{titlepage}

\section{Introduction}

It has been shown in \cite{ITA} that the instanton representation of Plebanski gravity exhibits a Hilbert space of harmonic oscillator coherent states.  In the present paper we will extend the construction of the states to include coupling to a massless spin 1/2 fermionic field.  The classical and quantum dynamics associated with fermions coupled to gravity is a subject of interest, since such coupling is relevant to the Standard Model.  Since the instanton representation uses the self-dual $SO(3,C)$ connection $A^a_i$, and this connection couples directly to the fermionic field via minimal coupling, then it follows that the connection can only couple to left handed fermions.  The purpose of this paper will be to quantize the coupled theory and construct its Hilbert space.  We will limit ourselves to minisuperspa for simplicity, by this we mean that all quantities are will be spatially homogeneous and depend only on time.  The organization of this paper is as follows.  Section 2 defines the coupled constraints and performs a reduction to minisuperspace.  Likewise in section 3 we reduce the canonical structure of the coupled theory to minisuperspace.  Due to the configuration of the connection used and some properties unique to fermions in the instanton representation, the canonical structure conspires to produce globally holonomic coordinates on configuration space.  This enables the quantization and subsequent construction of the Hilbert space, which we carry out in section 4.  In section 5 we write down the Hamilton's equations of motion, verifying consistency with the gravitational state labels previously defined.  Index conventions for this paper use Latin symbols $a,b,c,\dots$ from the beginning part of the Latin alphabet to denote internal indices, and symbols $i,j,k,\dots$ from the middle to denote spatial indices.  The mass dimensions for various quantities of interest are $[Q_d]=3$ and $[A^a_i]=1$, where $Q_d$ is the fermionic $SO(3,C)$ charge as well as $[\Psi_{ae}]=2$, where $\Psi_{ae}$ is the CDJ matrix.

\newpage

\section{Reduction of the constraints to minisuperspace}

\noindent
The 3+1 decomposition of the action for GR in Ashtekar variables coupled to a massless left-handed spin ${1 \over 2}$ fermionic field $\psi^A$ is given by \cite{ASHMATTER}

\begin{eqnarray}
\label{CANO}
S=\int{dt}\int_{\Sigma}d^3x{1 \over G}\widetilde{\sigma}^i_a\dot{A}^a_i+\pi_A\dot{\psi}^A\nonumber\\
-{N}(\hbox{det}\widetilde{\sigma})^{-1/2}\bigl({1 \over G}H_{grav}+H_{Ferm}\bigr)\nonumber\\
-N^i\bigl({1 \over G}(H_{grav})_i+(H_{Ferm})_i\bigr)-A^a_0\bigl({1 \over G}G_a+Q_a\bigr),
\end{eqnarray}

\noindent
where $(A^a_i,\widetilde{\sigma}^i_a)$ are the Ashtekar connection and the densitized triad, and $(\psi^A,\pi_A)$ are the left-handed fermionic field and its conjugate momentum.  The initial value constraints are smeared by auxilliary fields $(N,N^i,A^a_0)$, respectively the lapse function, 
shift vector and $SU(2)_{-}$ rotation angle.  The gravitational contribution to the Hamiltonian constraint $H_{grav}$ is given by

\begin{eqnarray}
\label{CANOI}
H_{grav}={\Lambda \over 3}\epsilon_{ijk}\epsilon^{abc}\widetilde{\sigma}^i_a\widetilde{\sigma}^j_b\widetilde{\sigma}^k_c
+\epsilon_{ijk}\epsilon^{abc}\widetilde{\sigma}^i_a\widetilde{\sigma}^j_bB^k_c,
\end{eqnarray}

\noindent
where $\Lambda$ is the cosmological constant.  The matter contribution to the Hamiltonian constraint is given by \cite{ASHMATTER}

\begin{eqnarray}
\label{CANOI1}
H_{Ferm}=\pi_A(\tau_a)^A_B\widetilde{\sigma}^i_aD_i\psi^B
\end{eqnarray}

\noindent
where $(\tau_a)^A_B$ are the Pauli matrices and $D_i$ is the gauge covariant derivative in the fundamental representation of left-handed $SU(2)_{-}$.  The diffeomorphism constraint is given by

\begin{eqnarray}
\label{CONOOO}
H_i={1 \over G}\epsilon_{ijk}\widetilde{\sigma}^j_aB^k_a+\pi_A(D_i\psi)^A=0,
\end{eqnarray}

\noindent
which in the full theory possesses a matter contribution containing spatial gradients.  The Gauss' law constraint is given by

\begin{eqnarray}
\label{GOOOSE}
{1 \over G}D_i\widetilde{\sigma}^i_a+Q_a=0,
\end{eqnarray}

\noindent
where $(D_i)^a_c=\delta^{ac}\partial_i+f^{abc}A^b_j$ is the $SU(2)_{-}$ covariant derivative with structure constants $f_{abc}$ and $Q_a$ is the fermionic $SU(2)_{-}$ charge, given by

\begin{eqnarray}
\label{CHARGEIT}
Q_a=\pi_A(\tau_a\psi)^A.
\end{eqnarray}

\noindent
We will now perform a reduction to minisuperspace,\footnote{Our definition of minisuperspace is that all dynamical variables are spatially constant, which is obtained by setting all spatial gradients of the variables
to zero.  We do not use \cite{KODAMA1}, which entails the introduction of Bianchi groups in the definition of minisuperspace.} in conjunction with transforming the gravitational variables of the starting action (\ref{CANO}) into the instanton representation of Plebanski gravity.  We will perform this transformation using the CDJ Ansatz

\begin{eqnarray}
\label{CANO1}
\widetilde{\sigma}^i_a=\Psi_{ae}B^i_e,
\end{eqnarray}

\noindent
where $\Psi_{ae}$ is the CDJ matrix, which takes its values in $SU(2)_{-}\otimes{SU}(2)_{-}$.  The Ashtekar magnetic field $B^i_a$ is related to the connection $A^a_i$ via

\begin{eqnarray}
\label{EIGEN2}
B^i_a=\epsilon^{ijk}\partial_jA^a_k+{1 \over 2}\epsilon^{ijk}f_{abc}A^b_jA^c_k=\epsilon^{ijk}\partial_jA^a_k+(\hbox{det}A)(A^{-1})^a_i,
\end{eqnarray}

\noindent
where we have used the fact that the structure constants $f_{abc}=\epsilon_{abc}$ for $SU(2)_{-}$, the gauge group for the self-dual Ashtekar variables, are numerically the same as the 3-D epsilon symbol.  For minisuperspace as defined, we set the spatial gradients in (\ref{EIGEN2}) to zero obtaining 

\begin{eqnarray}
\label{EIGEN3}
B^i_a=(\hbox{det}A)(A^{-1})^a_i;~~\hbox{det}B=(\hbox{det}A)^2;~~C_{ae}=A^a_iB^i_e=\delta_{ae}(\hbox{det}A).
\end{eqnarray}

\noindent
We will assume for the purposes of this paper that $(\hbox{det}A)\neq{0}$.

\subsection{Reduction of the kinematic constraints}

The kinematic initial value constraints are the Gauss' law and the diffeomorphism constraints $(G_a,H_i)$.  Substituting (\ref{CANO1}) into (\ref{CONOOO}), we obtain for the diffeomorphism constraint in the presence of fermionic fields the following

\begin{eqnarray}
\label{DIFFEOM}
\epsilon_{ijk}B^j_aB^k_e\Psi_{ae}+G(H_{Ferm})_i=0.
\end{eqnarray}

\noindent
For nondegenerate $B^i_a$, which we assume in this paper, the antisymmetric combination of magnetic fields can be replaced with their inverse by using the properties of determinants 
of 3 by 3 matrices.  Defining $\psi_d=\epsilon_{dae}\Psi_{ae}$ as the antisymmetric part of the CDJ matrix, (\ref{DIFFEOM}) can be written as

\begin{eqnarray}
\label{DIFFEOM1111}
(\hbox{det}B)(B^{-1})^d_i\psi_d+G(H_{Ferm})_i=0,
\end{eqnarray}

\noindent
which enables us to solve for $\psi_d$ as

\begin{eqnarray}
\label{DIFFEOM1}
\psi_d=-G(\hbox{det}B)^{-1}B^i_d(H_{Ferm})_i.
\end{eqnarray}

\noindent
The matter contribution to the diffeomorphism constraint $(H_{Ferm})_i$ can in turn be written as

\begin{eqnarray}
\label{DIFFEOM12}
(H_{Ferm})_i=\pi_A(D_i\psi)^A=\pi_A(\partial_i\psi)^A+A^a_i(\tau_a)^A_B\pi_A\psi^B
=\pi_A\partial_i\psi^A+A^a_iQ_a,
\end{eqnarray}

\noindent
Setting $\pi_A\partial_i\psi^A=0$ for minisuperspace and substituting (\ref{DIFFEOM12}) into (\ref{DIFFEOM1}), we have

\begin{eqnarray}
\label{DIFFEOM2}
\psi_d=-G(\hbox{det}B)^{-1}B^i_dA^a_iQ_a=-G(\hbox{det}B)^{-1}C_{da}Q_a=-G(\hbox{det}A)^{-1}Q_d
\end{eqnarray}

\noindent
where we have also used $C_{ae}=\delta_{ae}(\hbox{det}A)$ from (\ref{EIGEN3}) as well as $C^T_{ad}=C_{da}$.\par
\indent
To obtain the Gauss' law constraint in the new variables we substitute the CDJ Ansatz (\ref{CANO1}) into (\ref{GOOOSE}), yielding

\begin{eqnarray}
\label{GAUSS1}
D_i\widetilde{\sigma}^i_a=
\partial_i(\Psi_{ae}B^i_a)+f_{abc}A^b_iB^i_e\Psi_{ae}=-GQ_a.
\end{eqnarray}

\noindent
Setting $\partial_i(\Psi_{ae}B^i_a)=0$ for minisuperspace and including the matter contribution, then this reduces to

\begin{eqnarray}
\label{FFEE1}
f_{abc}A^b_iB^i_e\Psi_{ae}=f_{aec}\Psi_{ae}(\hbox{det}A)=-GQ_c
\end{eqnarray}

\noindent
where we have used (\ref{EIGEN3}).  Recalling the definition $\psi_d=\epsilon_{dae}\Psi_{ae}$ for the antisymmetric part of $\Psi_{ae}$, (\ref{FFEE1}) yields  

\begin{eqnarray}
\label{FFEE2}
\psi_d=-G(\hbox{det}A)^{-1}Q_d.
\end{eqnarray}

\noindent
Comparison of (\ref{FFEE2}) with (\ref{DIFFEOM2}) shows that in minisuperspace, the Gauss' law and the diffeomorphism constraints are redundant.  The result is that in minisuperspace the kinematic constraints combined can only reduce the degrees of freedom of the CDJ matrix by three and not six.\par
\indent

\subsection{Reduction of the Hamiltonian constraint}

\noindent
We are now ready to transfrom the Hamiltonian constraint into the instanton representation and reduce it to minisuperspace.  Substituting (\ref{CANO1}) into (\ref{CANOI}), we have  

\begin{eqnarray}
\label{HAMILTONIAN}
(\hbox{det}B)\bigl({1 \over 2}Var\Psi+\Lambda\hbox{det}\Psi\bigr)+GH_{Ferm}=0
\end{eqnarray}

\noindent
where we have defined $Var\Psi=(\hbox{tr}\Psi)^2-\hbox{tr}\Psi^2$.  Using (\ref{EIGEN3}) for minisuperspace, this reduces to

\begin{eqnarray}
\label{HAMILTONIAN1}
(\hbox{det}A)^2\bigl({1 \over 2}Var\Psi+\Lambda\hbox{det}\Psi\bigr)+GH_{Ferm}=0.
\end{eqnarray}

\noindent
From (\ref{CANOI1}) we have the fermionic contribution

\begin{eqnarray}
\label{FERMI}
H_{Ferm}=\pi_A(\tau_a)^A_B\widetilde{\sigma}^i_aD_i\psi^B.
\end{eqnarray}

\noindent
Substituting (\ref{CANO1}) into (\ref{FERMI}) we have

\begin{eqnarray}
\label{FERMI1}
H_{Ferm}=\pi_A(\tau_a)^A_B\widetilde{\sigma}^i_aD_i\psi^B=\pi_A(\tau_a)^A_B\Psi_{ae}B^i_e(D_i\psi)^B\nonumber\\
=\Psi_{ae}\pi_A(\tau_a)^A_BB^i_e\bigl(\partial_i\psi^B+A^f_i(\tau_f)^B_C\psi^C\bigr),
\end{eqnarray}

\noindent
where the covariant derivative acts on left handed spinors.  Continuing the expansion we have

\begin{eqnarray}
\label{FERMI2}
H_{Ferm}=\Psi_{ae}\pi_A(\tau_a)^A_B\textbf{v}_e\{\psi^B\}+\Psi_{ae}C^T_{fe}\pi_A(\tau_a\tau_f)^A_C\psi^C,
\end{eqnarray}

\noindent
where $C_{fe}=A^f_iB^i_e$, and we have defined the vector field $\textbf{v}_e=B^i_a\partial_i$.  To evaluate the product of Pauli matrices we make use of the identity

\begin{eqnarray}
\label{FERMI3}
(\tau_a\tau_f)^A_C=\{\tau_a,\tau_f\}^A_C+[\tau_a,\tau_f]^A_C
=\delta_{af}\delta^A_C+i\epsilon_{afd}(\tau_d)^A_C.
\end{eqnarray}

\noindent
Putting (\ref{FERMI3}) into (\ref{FERMI2}) we have

\begin{eqnarray}
\label{FERMI4}
H_{Ferm}=\Psi_{ae}\pi_A(\tau_a)^A_B\textbf{v}_e\{\psi^B\}+C^T_{fe}\Psi_{ae}\pi_A\bigl(\delta_{af}\delta^A_C+i\epsilon_{afd}(\tau_d)^A_C\bigr)\psi^C.
\end{eqnarray}

\noindent
Making the definitions

\begin{eqnarray}
\label{FERMI5}
Q=\pi_A\psi^A;~~Q_d=\pi_A(\tau_d\psi)^A,
\end{eqnarray}

\noindent
and setting spatial gradients to zero for minisuperspace we have

\begin{eqnarray}
\label{FERMI6}
H_{Ferm}=C_{fe}\bigl(\Psi_{fe}Q+i\epsilon_{afd}\Psi_{ae}Q_d\bigr).
\end{eqnarray}

\noindent
Upon use of $C_{be}=\delta_{be}(\hbox{det}A)=C^T_{eb}$ and (\ref{FFEE1}) we obtain

\begin{eqnarray}
\label{FERMI7}
H_{Ferm}=(\hbox{det}A)(\hbox{tr}\Psi)Q-iGQ_dQ_d.
\end{eqnarray}

\noindent
Note the cancellation of the factor of $(\hbox{det}A)$ in the second term of (\ref{FERMI7}).  The matter charges are given by

\begin{displaymath}
Q_1=
\left(\begin{array}{cc}
\pi_1 & \pi_2
\end{array}\right)
\left(\begin{array}{cc}
0 & 1\\
1 & 0\\
\end{array}\right)
\left(\begin{array}{c}
\psi^1\\
\psi^2\\
\end{array}\right)
=\pi_1\psi^2+\pi_2\psi^1
\end{displaymath}

\begin{displaymath}
Q_2=
\left(\begin{array}{cc}
\pi_1 & \pi_2
\end{array}\right)
\left(\begin{array}{cc}
0 & -i\\
i & 0\\
\end{array}\right)
\left(\begin{array}{c}
\psi^1\\
\psi^2\\
\end{array}\right)
=i(-\pi_1\psi^2+\pi_2\psi^1)
\end{displaymath}

\begin{displaymath}
Q_3=
\left(\begin{array}{cc}
\pi_1 & \pi_2
\end{array}\right)
\left(\begin{array}{cc}
1 & 0\\
0 & -1\\
\end{array}\right)
\left(\begin{array}{c}
\psi^1\\
\psi^2\\
\end{array}\right)
=\pi_1\psi^1-\pi_2\psi^2,
\end{displaymath}

\noindent
where we have used the normal-ordering convention of momenta to the left of the coordinates for the fermionic fields.  Also, we will use the anticommutative relations for Grassman numbers $\xi$, $\psi$ that

\begin{eqnarray}
\label{LEFT8}
\chi\psi=-\psi\chi.
\end{eqnarray}

\noindent
We also have 

\begin{displaymath}
Q=
\left(\begin{array}{cc}
\pi_1 & \pi_2
\end{array}\right)
\left(\begin{array}{cc}
1 & 0\\
0 & 1\\
\end{array}\right)
\left(\begin{array}{c}
\psi^1\\
\psi^2\\
\end{array}\right)
=\pi_1\psi^1+\pi_2\psi^2.
\end{displaymath}

\noindent
From these relations and conventions one sees that the square of all components of the charges $Q,Q_d$ are the same

\begin{eqnarray}
\label{LEFT9}
Q_1Q_1=Q_2Q_2=Q_3Q_3=-2(\pi_1\psi^1)(\pi_2\psi^2)=-Q^2.
\end{eqnarray}

\noindent
Hence using (\ref{LEFT9}) in (\ref{FERMI7}), we see that the fermionic contribution to the Hamiltonian constraint, multiplying a factor of $G$, is given by

\begin{eqnarray}
\label{LEFT91}
GH_{Ferm}=(\hbox{det}A)(\hbox{tr}\Psi)(GQ)+3i(GQ)^2.
\end{eqnarray}

\par
\indent
The CDJ matrix can be parametrized using its irreducible parts under internal rotations

\begin{eqnarray}
\label{PARAMETERIZED}
\Psi_{ae}=\delta_{ae}\varphi_e+S_{ae}+\epsilon_{aed}\psi_d,
\end{eqnarray}

\noindent
where $S_{ae}$ are the off-diagonal symmetric (shear) elements.  Recall that the diffeomorphism and the Gauss' law constraints fix only three out of six elements of $\Psi_{ae}$, namely $\psi_d$.  Hence after implementation of the kinematic constraints this leaves remaining three freely specifiable elements $S_{ae}$.  For this paper we will choose $S_{ae}=0$ for simplicity, leaving 

\begin{displaymath}
\Psi_{ae}=\delta_{ae}\varphi_e+\epsilon_{aed}\psi_d=
\left(\begin{array}{ccc}
\varphi_1 & \psi_3 & -\psi_2\\
-\psi_3 & \varphi_2 & \psi_1\\
\psi_2 & -\psi_1 & \varphi_3\\
\end{array}\right)
.
\end{displaymath}

\noindent
Hence upon implementation of the kinematic constraints, (\ref{HAMILTONIAN}) amounts to a condition on the invariants of $\Psi_{ae}$, which can be written in terms of the diagonal elements $\varphi_f$.  The invariants are given by

\begin{eqnarray}
\label{INVARI}
Var\Psi=(\hbox{tr}\Psi)^2-\hbox{tr}\Psi^2=2\Bigl(\varphi_1\varphi_2+\varphi_2\varphi_3+\varphi_3\varphi_1+(\psi_1)^2+(\psi_2)^2+(\psi_3)^2\Bigr)
\end{eqnarray}

\noindent
for the variance, and

\begin{eqnarray}
\label{INVARI1}
\hbox{det}\Psi=\varphi_1\varphi_2\varphi_3+\varphi_1(\psi_1)^2+\varphi_2(\psi_2)^2+\varphi_3(\psi_3)^2
\end{eqnarray}

\noindent
for the determinant.  Substituting (\ref{FFEE2}) into (\ref{INVARI}) and (\ref{INVARI1}), we obtain the following 

\begin{eqnarray}
\label{INVARI2}
{1 \over 2}Var\Psi=\varphi_1\varphi_2+\varphi_2\varphi_3+\varphi_3\varphi_1-3(GQ)^2(\hbox{det}A)^{-2};\nonumber\\
\hbox{det}\Psi=\varphi_1\varphi_2\varphi_3-(\varphi_1+\varphi_2+\varphi_3)(GQ)^2(\hbox{det}A)^{-2}
\end{eqnarray}

\noindent
where we have used (\ref{LEFT9}).  Substituting (\ref{INVARI2}) and (\ref{LEFT91}) into (\ref{HAMILTONIAN1}) we obtain

\begin{eqnarray}
\label{HAMEELTO}
(\hbox{det}A)^2\biggl[\varphi_1\varphi_2+\varphi_2\varphi_3+\varphi_3\varphi_1-3(GQ)^2(\hbox{det}A)^{-2}\nonumber\\
+\Lambda\bigl(\varphi_1\varphi_2\varphi_3-(\varphi_1+\varphi_2+\varphi_3)(GQ)^2(\hbox{det}A)^{-2}\bigr)\biggr]\nonumber\\
+(\varphi_1+\varphi_2+\varphi_3)(\hbox{det}A)(GQ)+3i(GQ)^2=0.
\end{eqnarray}

\noindent
Next, we define densitized gravitational momentum space variables by

\begin{eqnarray}
\label{DENSITIZED}
\widetilde{\varphi}_f=\varphi_f(\hbox{det}A)
\end{eqnarray}

\noindent
for $f=1,2,3$, with mass dimension $[\widetilde{\varphi}]=1$.  Then the Hamiltonian constraint can be written in the form

\begin{eqnarray}
\label{DENSITIZED1}
\biggl[\bigl(\widetilde{\varphi}_1\widetilde{\varphi}_2+\widetilde{\varphi}_2\widetilde{\varphi}_3+\widetilde{\varphi}_3\widetilde{\varphi}_1\bigr)
+(\widetilde{\varphi}_1+\widetilde{\varphi}_2+\widetilde{\varphi}_3)(GQ)+3(i-1)(GQ)^2\biggr]\nonumber\\
+\Lambda(\hbox{det}A)^{-1}\Bigl(\widetilde{\varphi}_1\widetilde{\varphi}_2\widetilde{\varphi}_3-(\widetilde{\varphi}_1+\widetilde{\varphi}_2+\widetilde{\varphi}_3)(GQ)^2\Bigr)=0.
\end{eqnarray}

\noindent
We will now determine the canonical structure of the theory in preparation for quantization.  Since the Hamiltonian constraint depends only on the invariants of $\Psi_{ae}$ and we have reduced this to the 
diagonal elements $\varphi_f$, then it follows without loss of generality that we can identify $\varphi_f$ with the eigenvalues of $\Psi_{(ae)}$, the symmetric part of $\Psi_{ae}$.

\newpage

\section{Reduction of the canonical structure}

\noindent
We have thus far performed a transformation of the gravitational variables into the instanton representation of Plebanski gravity, in conjunction with a reduction into minisuperspace.  To determine the canonical structure of the theory we must do the same for the canonical one form

\begin{eqnarray}
\label{TODETERM}
\boldsymbol{\theta}=\int_{\Sigma}d^3x\Bigl({1 \over G}\widetilde{\sigma}^i_a\delta{A}^a_i+\pi_A\delta\psi^A\Bigr).
\end{eqnarray}

\noindent
Substituting (\ref{CANO1}) in conjunction with (\ref{EIGEN3}) into (\ref{TODETERM}), we obtain for the gravitational contribution that

\begin{eqnarray}
\label{ONEFORM}
\boldsymbol{\theta}_{Grav}={1 \over G}\int_{\Sigma}d^3x\Psi_{ae}B^i_e\delta{A}^a_i
={{l^3} \over G}\Psi_{ae}(\hbox{det}A)(A^{-1})^i_e\delta{A}^a_i,
\end{eqnarray}

\noindent
where $l$ is the characteristic length scale of the universe, which arises from integration of the spatially homogeneous variables over 3-space $\Sigma$.  Equation (\ref{ONEFORM}) can be written as

\begin{eqnarray}
\label{WRITTENAS}
\boldsymbol{\theta}_{Grav}={{l^3} \over G}\Psi_{(ae)}(\hbox{det}A)(A^{-1})^i_e\delta{A}^a_i+{{l^3} \over G}\Psi_{[ae]}(\hbox{det}A)(A^{-1})^i_e\delta{A}^a_i,
\end{eqnarray}

\noindent
which separates the CDJ matrix into its symmetric and its antisymmetric parts.  We will implement the diffeomorphism and Gauss' law constraints at the level of (\ref{WRITTENAS}).  There are two parts, one part symmetric in $\Psi_{ae}$ and the other part antisymmetric.  We will first focus on the antisymmetric part, which will require use of (\ref{FFEE2}), requoted here

\begin{eqnarray}
\label{THISCAN}
\epsilon_{dae}\Psi_{ae}=-GQ_d(\hbox{det}A)^{-1}.
\end{eqnarray}

\noindent
Substituting (\ref{THISCAN}) into the second term on the right hand side of (\ref{WRITTENAS}), we have

\begin{eqnarray}
\label{THISCAN1}
{{l^3} \over G}\bigl({1 \over 2}\epsilon_{aed}\epsilon_{dbf}\Psi_{bf}\bigr)(\hbox{det}A)(A^{-1})^i_e\delta{A}^a_i
=-{{l^3} \over 2}\epsilon_{dae}Q_d(A^{-1})^i_e\delta{A}^a_i.
\end{eqnarray}

\noindent
There is no configuration space coordinate corresponding to $(A^{-1})^i_{[e}\delta{A}^a_{i]}$, since it is not an exact functional one form, which presents an obstruction to quantization.  However, if we restrict attention to a diagonal connection $A^a_i=\delta^a_iA^a_a$, then (\ref{THISCAN1}) reduces to

\begin{eqnarray}
\label{THISCAN2}
-{{l^3} \over 2}\epsilon_{dae}Q_d\delta^i_e\Bigl({1 \over {A^a_e}}\Bigr)\delta^a_i\delta{A}^a_a
=-{{l^3} \over 2}\delta_{ae}\epsilon_{aed}Q_d\Bigl({{\delta{A}^a_a} \over {A^e_e}}\Bigr)=0,
\end{eqnarray}

\noindent
which vanishes on account of antisymmetry of $\epsilon_{aed}$.  Hence for a diagonal connection, the second term on the right hand side of (\ref{WRITTENAS}) vanishes, leaving the first term evaluated on a diagonal connection.  It then remains to show that the $\Psi_{(ae)}$ contribution is in fact quantizable.\par
\indent
Define densitized gravitational momentum space variables $\widetilde{\Psi}_{ae}=\Psi_{ae}(\hbox{det}A)$.  Without loss of generality we can choose a diagonal CDJ matrix and identify the diagonal elements with the 
eigenvalues $\varphi_f$, simplifying (\ref{WRITTENAS}) to

\begin{eqnarray}
\label{ONEFORM1}
\boldsymbol{\theta}_{Grav}={{l^3} \over G}\widetilde{\Psi}_{aa}(A^{-1})^i_a\delta{A}^a_i\equiv{{l^3} \over G}\widetilde{\varphi}_a(A^{-1})^i_a\delta{A}^a_i.
\end{eqnarray}

\noindent
For a diagonal connection $A^a_i=\delta^a_iA^a_a$, the following relations can be written

\begin{eqnarray}
\label{ONEFORM2}
(A^{-1})^i_1\delta{A}^1_i={{\delta{A}^1_1} \over {A^1_1}};~~
(A^{-1})^i_2\delta{A}^2_i={{\delta{A}^2_2} \over {A^2_2}};~~
(A^{-1})^i_3\delta{A}^3_i={{\delta{A}^3_3} \over {A^3_3}}.
\end{eqnarray}

\noindent
Equation (\ref{ONEFORM}) implies the existence of dimensionless holonomic configuration space coordinates $(X,Y,T)$ with

\begin{eqnarray}
\label{PRONTO}
X=\hbox{ln}\Bigl({{A^1_1} \over {a_0}}\Bigr);~~Y=\hbox{ln}\Bigl({{A^2_2} \over {a_0}}\Bigr);~~
T=\hbox{ln}\Bigl({{A^1_1A^2_2A^3_3} \over {a_0^3}}\Bigr),
\end{eqnarray}

\noindent
where $a_0$ is a mass scale.  The matter contribution to the canonical one form is given by

\begin{eqnarray}
\label{ONEFORM4}
\boldsymbol{\theta}_{Ferm}=\int_{\Sigma}d^3x\pi_A\delta\psi^A=l^3\pi_A\delta{\psi}^A.
\end{eqnarray}

\noindent
In order that this be dimensionless this implies that the mass dimensions of the matter fields are $[\pi_A]=[\psi^A]={3 \over 2}$.\par
\indent
Hence the total symplectic two form is given by

\begin{eqnarray}
\label{HENCETHE}
\boldsymbol{\omega}={{l^3} \over G}{\delta\varphi_a}\wedge{\delta(\hbox{ln}\Bigl({{A^a_a} \over {a_0}}\Bigr))}+l^3{\delta\pi_A}\wedge{\delta\psi^A}\nonumber\\
=l^3\delta\Bigl({1 \over G}\varphi_a{\delta(\hbox{ln}\Bigl({{A^a_a} \over {a_0}}\Bigr))}+\pi_A\delta{\psi}^A\Bigr).
\end{eqnarray}

\noindent
Defining ${{l^3} \over G}\equiv{1 \over \mu}$, where $[\mu]=1$, we see that (\ref{HENCETHE}) implies the following nonvanishing Poisson brackets

\begin{eqnarray}
\label{ONEFORM3}
\{\hbox{ln}\Bigl({{A^f_f} \over {a_0}}\Bigr),\widetilde{\varphi}_g\}=\mu\delta^f_g;~~\{\psi^A,\pi_B\}={1 \over {l^3}}\delta^A_B,
\end{eqnarray}

\noindent
which take into account the anticommutative nature of the fermionic variables.  Prior to quantization we will perform the following change to dimensionless gravitational momentum space variables

\begin{eqnarray}
\label{PRIORTO}
a_3={1 \over \mu}\widetilde{\varphi}_3;~~a_1={1 \over \mu}(\widetilde{\varphi}_1-\widetilde{\varphi}_3);~~a_2={1 \over \mu}(\widetilde{\varphi}_2-\widetilde{\varphi}_3),
\end{eqnarray}

\noindent
Comparison with (\ref{ONEFORM3}) fixes the canonically conjugate configuration variables to $a_1$, $a_2$ and $a_3$ respectively as $X$, $Y$ and $T$.  Upon quantization the gravitational momentum space variables will take on the following Schr\"odinger representation

\begin{eqnarray}
\label{ONEFORM8}
a_1\longrightarrow{\partial \over {\partial{X}}};~~a_2\longrightarrow{\partial \over {\partial{Y}}};~~
a_3\longrightarrow{\partial \over {\partial{T}}}.
\end{eqnarray}

\noindent
We will identify $a_f$ as the annihilation operators for three uncoupled simple harmonic oscillators.  Then from equation (\ref{ONEFORM8}) when one makes the following identifications upon quantization

\begin{eqnarray}
\label{ONEFORM9}
X\longrightarrow{a}_1^{\dagger};~~Y\longrightarrow{a}_2^{\dagger};~~T\longrightarrow{a}_3^{\dagger},
\end{eqnarray}

\noindent
this fixes the representation space for the gravitational variables according to the Bargmann representation on holomorphic functions.\par
\par
\indent
Having identified globally holonomic coordinates for the gravitational and the matter variables, we can now write a symplectic two form which is an exact variation of the corresponding canonical one form

\begin{eqnarray}
\label{ONEFORM7}
\boldsymbol{\omega}={\delta{a}_f}\wedge{\delta{a}^{*}_f}+l^3{\delta\pi_A}\wedge{\delta\psi^A}=\delta\boldsymbol{\theta},
\end{eqnarray}

\noindent
where $\boldsymbol{\theta}=\boldsymbol{\theta}_{Ferm}+\boldsymbol{\theta}_{Grav}$.  To obtain the Hamiltonian constraint in the oscillator variables we must first invert (\ref{PRIORTO})

\begin{eqnarray}
\label{MUSTINVERT}
\widetilde{\varphi}_1=\mu(a_1+a_3);~~\varphi_2=\mu(a_2+a_3);~~\varphi_3=\mu{a}_3,
\end{eqnarray}

\noindent
and substitute (\ref{MUSTINVERT}) into (\ref{DENSITIZED1}).  This yields a Hamiltonian constraint of 

\begin{eqnarray}
\label{PRIORTO1}
3\mu^2\bigl(a_3a_3+{2 \over 3}(a_1+a_2)a_3+{1 \over 3}a_1a_2\bigr)+3\mu(GQ)\bigl(a_3+{1 \over 3}(a_1+a_2)\bigr)\nonumber\\
+3(i-1)(GQ)^2
+{\Lambda \over {a_0^3}}e^{-a_3^{*}}\Bigl[\mu^3a_3(a_3+a_1)(a_3+a_2)-3\mu(GQ)^2\bigl(a_3+{1 \over 3}(a_1+a_2)\bigr)\Bigr]=0.
\end{eqnarray}

\noindent

\section{Quantization and construction of the Hilbert space}

\noindent
The gravitational variables $a_f$ upon quantization will become promoted to operators satisfying commutation relations  

\begin{eqnarray}
\label{PRIORTO2}
[\hat{a}_f,\hat{a}^{\dagger}_g]=\delta_{fg},
\end{eqnarray}

\noindent
where $\hat{a}_f^{\dagger}$ is the quantum analogue of $a^{*}_f$, the complex conjugate of $a_f$.  This implies that $\hat{a}_f$ and $\hat{a}^{\dagger}_f$ respectively are annihilation and creation operators for a triple of uncoupled simple harmonic oscillators.  Each harmonic oscillator comes equipped with a ground state $\bigl\vert{0}\bigr>$ such that

\begin{eqnarray}
\label{COMESWITH}
a_f\bigl\vert{0}\bigr>=0
\end{eqnarray}

\noindent
for $f=1,2,3$.  Define coherent states as eigenstates of the annihilation operators

\begin{eqnarray}
\label{PRIORTO3}
\hat{a}_1\bigl\vert\alpha\bigr>=\alpha\bigl\vert\alpha\bigr>;~~\hat{a}_2\bigl\vert\beta\bigr>=\beta\bigl\vert\beta\bigr>;~~\hat{a}_3\bigl\vert\lambda\bigr>=\lambda\bigl\vert\lambda\bigr>.
\end{eqnarray}

\noindent
Equation (\ref{PRIORTO3}) are also Perelomov coherent states in the sense that they correspond to displacement of the `vacuum' state $\bigl\vert{0}\bigr>$ into a complex two-dimensional space $(\alpha,\beta)\in{C}_2$, which plays the role of the coset manifold of the complexified Heisenberg group.  The displacement operators act via

\begin{eqnarray}
\label{PRIORTO31}
\bigl\vert\alpha\bigr>=e^{\alpha{a}_1^{\dagger}-\alpha^{*}a_1}\bigl\vert{0}\bigr>;~~
\bigl\vert\beta\bigr>=e^{\beta{a}_2^{\dagger}-\beta^{*}a_2}\bigl\vert{0}\bigr>;~~
\bigl\vert\lambda\bigr>=e^{\lambda{a}_3^{\dagger}}\bigl\vert{0}\bigr>.
\end{eqnarray}

\noindent
We have singled out $a_3$ from $a_1$ and $a_2$, which implies that we will not be performing a normalization with respect to $\bigl\vert\lambda\bigr>$.  This is because we will associate $a_3$ to a time variable on the gravitational phase space, and one does not normalize a wavefunction in time.  The states (\ref{PRIORTO31}) will form the basis of the gravitational part of the Hilbert space of the coupled system.\par
\indent
Since the matter variables $(\psi^A,\pi_A)$ are Grassman numbers, they will upon quantization satisfy equal-time anticommutation relations

\begin{eqnarray}
\label{PRIORTO21}
[\hat{\psi}^A,\hat{\pi}_B]_{+}=l^{-3}\delta^A_B.
\end{eqnarray}

\noindent
The matter operators have a Schr\"odinger representation

\begin{eqnarray}
\label{MATTER}
\hat{\pi}_A=l^{-3}{\partial \over {\partial\psi^A}};~~G\hat{Q}=l^{-3}\Bigl(\psi^1{\partial \over {\partial\psi^1}}+\psi^2{\partial \over {\partial\psi^2}}\Bigr).
\end{eqnarray}

\noindent

\noindent
For the matter part of the state, the most general left-handed fermionic state is polynomial in Grassman numbers

\begin{eqnarray}
\label{FERMIONICSTATE}
\Phi=c_0+c_1\psi^1+c_2\psi^2+c_{12}\psi^1\psi^2.
\end{eqnarray}

\noindent
The highest order constructible from 2-component left handed spinors is bilinear due to the anticommuting property of Grassman numbers.  Therefore there are four orthonormal basis states

\begin{eqnarray}
\label{MAXIMUM}
\bigl\vert\phi_0\bigr>\equiv{1};~~\bigl\vert\phi_1\bigr>\equiv\psi^1;~~\bigl\vert\phi_2\bigr>\equiv\psi^2;~~\bigl\vert\phi_3\bigr>\equiv\psi^1\psi^2,
\end{eqnarray}

\noindent
namely one empty state and one two-fermion state, and two single fermion states.  Note that each $\bigl\vert\phi_m\bigr>$ is an eigenstate of the matter charge operator

\begin{eqnarray}
\label{FERMIONICSTATE1}
G\hat{Q}\bigl\vert\phi_0\bigr>=0;~~
G\hat{Q}\bigl\vert\phi_1\bigr>=\bigl\vert\phi_1\bigr>;~~
G\hat{Q}\bigl\vert\phi_2\bigr>=\bigl\vert\phi_2\bigr>;~~
G\hat{Q}\bigl\vert\phi_3\bigr>=2\bigl\vert\phi_3\bigr>,
\end{eqnarray}

\noindent
where the fermion number is the eigenvalue, denoted by $k=(0,1,2)$.  The fermionic states are orthogonal

\begin{eqnarray}
\label{FERMORTHO}
\bigl<\phi_i\bigl\vert\phi_j\bigr>=\delta_{ij}.
\end{eqnarray}

\noindent
To quantize the Hamiltonian constraint we promote the dynamical variables to operators in (\ref{PRIORTO1}).  Then dividing by $3\mu^2$ yields the dimensionless equation

\begin{eqnarray}
\label{HAMILTONIANNN}
\biggl[\hat{a}_3\hat{a}_3+{2 \over 3}(\hat{a}_1+\hat{a}_2)\hat{a}_3+{1 \over 3}\hat{a}_1\hat{a}_2
+\Bigl({{G\hat{Q}} \over \mu}\Bigr)\bigl(\hat{a}_3+{1 \over 3}(\hat{a}_1+\hat{a}_2)\bigr)+(i-1)\Bigl({{G\hat{Q}} \over \mu}\Bigr)^2\nonumber\\
+\Bigl({{\mu\Lambda} \over {3a_0^3}}\Bigr)
\Bigl[\hat{a}_3(\hat{a}_3+\hat{a}_1)(\hat{a}_3+\hat{a}_2)-3\bigl(\hat{a}_3+{1 \over 3}(\hat{a}_1+\hat{a}_2)\bigr)\Bigl({{G\hat{Q}} \over \mu}\Bigr)^2\Bigr]e^{-\hat{a}_3^{\dagger}}\biggr]\boldsymbol{\psi}=0.
\end{eqnarray}

\noindent
Note that (\ref{PRIORTO31}) and (\ref{MAXIMUM}) are eigenstates of the Hamiltonian constraint operator, hence we can find $\boldsymbol{\psi}\in{Ker}\{\hat{H}\}$ by considering states of the form 

\begin{eqnarray}
\label{FERMIONICSTATE2}
\boldsymbol{\psi}=\bigl\vert\alpha,\beta\bigr>\otimes\bigl\vert\phi_i\bigr>\otimes\chi(T)
\end{eqnarray}

\noindent
where $\bigl\vert\alpha,\beta\bigr>=\bigl\vert\alpha\bigr>\otimes\bigl\vert\beta\bigr>$, and $\bigl\vert\phi\bigr>$ is one of the terms from (\ref{FERMIONICSTATE}), and $\chi$ is annihilated by $a_1$ and $a_2$, but not by $a_3$.   We will replace the action of $\hat{a}_1$, $\hat{a}_2$ and $G\hat{Q}$ in (\ref{HAMILTONIANNN}) on the state by their eigenvalues $\alpha$, $\beta$ and $k$, leaving intact the action of $\hat{a}_3$ and $\hat{a}_3^{\dagger}$ on $\chi$ since these will be associated to time variables.  The quantum Hamiltonian constraint then reduces to  

\begin{eqnarray}
\label{FERMIONICSTATE4}
\biggl[\hat{a}_3\hat{a}_3+{2 \over 3}(\alpha+\beta)\hat{a}_3+{1 \over 3}\alpha\beta
+k\bigl(\hat{a}_3+{1 \over 3}(\alpha+\beta)\bigr)+k^2(i-1)\nonumber\\
+\Bigl({{\mu\Lambda} \over {3a_0^3}}\Bigr)
\Bigl[\hat{a}_3(\hat{a}_3+\alpha)(\hat{a}_3+\beta)-3k^2\bigl(\hat{a}_3+{1 \over 3}(\alpha+\beta)\bigr)\Bigr]e^{-a_3^{\dagger}}\biggr]\boldsymbol{\psi}=0.
\end{eqnarray}

\noindent
We will define the following Schr\"odinger representation for $a_3$, since we have singled it out as special

\begin{eqnarray}
\label{FERMIONICSTATE5}
\hat{a}_3\chi(T)={\partial \over {\partial{T}}}\chi(T);~~\hat{a}_3^{\dagger}\chi(T)=T\chi(T),
\end{eqnarray}

\noindent
where $\chi(T)$ is in the holomorphic representation of $T$.  Define the following polynomials

\begin{eqnarray}
\label{FERMIONICSTATE6}
q^2+\bigl(k+{2 \over 3}(\alpha+\beta)\bigr)q+{1 \over 3}\alpha\beta+{k \over 3}(\alpha+\beta)+k^2(i-1)=(q+m_{-})(q+m_{+});\nonumber\\
q^3+(\alpha+\beta)q^2+(\alpha\beta-3k^2)q-k^2(\alpha+\beta)=(q+j_{-})(q+j_{0})(q+j_{+}),
\end{eqnarray}

\noindent
where $m_{\pm}$ and $j_{-}$, $j_{0}$ and $j_{+}$ are the corresponding roots.  Note that these roots are labelled by $\alpha$, $\beta$ and $k$, the same quantities that label the gravitational and matter basis wavefunctions.  Also define the dimensionless constant $r$, given by

\begin{eqnarray}
\label{FERMIONICSTATE7}
r=\Bigl({{\mu\Lambda} \over {3a_0^3}}\Bigr).
\end{eqnarray}

\noindent
Then the quantum Hamiltonian constraint is given by

\begin{eqnarray}
\label{ISGIVNEBY}
\Bigl[(\hat{a}_3+m_{-})(\hat{a}_3+m_{+})+r(\hat{a}_3+j_{-})(\hat{a}_3+j_{0})(\hat{a}_3+j_{+})e^{\hat{a}_3^{\dagger}}\Bigr]\boldsymbol{\psi}=0.
\end{eqnarray}

\noindent
Additionally, define the dimensionless variable $z$, where

\begin{eqnarray}
\label{FERMIONICSTATE8}
{1 \over r}e^{a_3^{\dagger}}={1 \over r}e^T\equiv{z}\longrightarrow\hat{a}_3={\partial \over {\partial{T}}}=z{d \over {dz}}.
\end{eqnarray}

\noindent
Then the Hamiltonian constraint reduces to the following ordinary differential equation in $z$

\begin{eqnarray}
\label{FERMIONICSTATE9}
\biggl[\Bigl(z{d \over {dz}}+m_{-}\Bigr)\Bigl(z{d \over {dz}}+m_{+}\Bigr)z+\Bigl(z{d \over {dz}}+j_{-}\Bigr)\Bigl(z{d \over {dz}}+j_0\Bigr)\Bigl(z{d \over {dz}}+j_{+}\Bigr)\biggr]\chi(z)=0.
\end{eqnarray}

\noindent
Using the identity

\begin{eqnarray}
\label{FERMIONICSTATE10}
z{d \over {dz}}(zF)=z\Bigl(z{d \over {dz}}+1\Bigr)F,
\end{eqnarray}

\noindent
we can commute the variable $z$ to the left, resulting in the following diffferential equation

\begin{eqnarray}
\label{FERMIONICSTATE11}
\biggl[z\Bigl(z{d \over {dz}}+m_{-}+1\Bigr)\Bigl(z{d \over {dz}}+m_{+}+1\Bigr)+\Bigl(z{d \over {dz}}+j_{-}\Bigr)\Bigl(z{d \over {dz}}+j_0\Bigr)\Bigl(z{d \over {dz}}+j_{+}\Bigr)\biggr]\chi(z)=0.
\end{eqnarray}

\noindent
Equation (\ref{FERMIONICSTATE11}) is the hypergeometric differential equation, with solution

\begin{eqnarray}
\label{FERMIONICSTATE12}
\chi(z)={_3F_3}(1,m_{-}+1,m_{+}+1;j_{-},j_{0},j_{+};z).
\end{eqnarray}

\noindent
The quantum state then is given by

\begin{eqnarray}
\label{THESTATEIS}
\boldsymbol{\psi}=\boldsymbol{\psi}_{\alpha,\beta,k}(T)=\bigl\vert\alpha,\beta\bigr>\otimes\bigl\vert\phi_k\bigr>\otimes\chi(T),
\end{eqnarray}

\noindent
which $\chi$ given by (\ref{FERMIONICSTATE12}).  The wavefunctions corresponding to (\ref{THESTATEIS}) are convergent, owing to the convergent properties of the hypergeometric function $_3F_3$, and are also normalizable.

\newpage

\section{Classical dynamics}

Having found the Hilbert space solving the quantum Hamiltonian constraint, we will now check for consistency with the classical dynamics.  In particular, we would like to see if the gravitational coherent states are preserved under fermionic coupling during the time evolution.  To do this we will transform the gravitational variables back into the Schr\"odinger representation and analyse the Hamiltonian dynamics.  The Hamiltonian can be written in the following form

\begin{eqnarray}
\label{DYNAMICS}
\boldsymbol{H}={N \over {\sqrt{\hbox{det}\widetilde{\sigma}}}}\Bigl({1 \over G}(\hbox{det}B)\bigl({1 \over 2}Var\Psi+\Lambda\hbox{det}\Psi\bigr)+H_{Ferm}\Bigr)\nonumber\\
=N\sqrt{\hbox{det}\widetilde{\sigma}}\Bigl({1 \over {(\hbox{det}\Psi)(\hbox{det}B)}}\Bigr)
\Bigl({1 \over G}(\hbox{det}B)\bigl({1 \over 2}Var\Psi+\Lambda\hbox{det}\Psi\bigr)+H_{Ferm}\Bigr),
\end{eqnarray}

\noindent
where we have used the CDJ Ansatz.  The next step is to reduce (\ref{DYNAMICS}) to minisuperspace using $\hbox{det}B=(\hbox{det}A)^2$, followed by densitization of the CDJ matrix

\begin{eqnarray}
\label{DYNAMICS1}
\widetilde{\Psi}_{ae}=(\hbox{det}A)\Psi_{ae}.
\end{eqnarray}

\noindent
The result of applying these steps to (\ref{DYNAMICS}) yields

\begin{eqnarray}
\label{DYNAMICS2}
\boldsymbol{H}
=N(\hbox{det}A)\sqrt{\hbox{det}\Psi}\Bigl({1 \over G}\bigl(\Lambda+\hbox{tr}\Psi^{-1}\bigr)+{1 \over {\hbox{det}\Psi}}(\hbox{det}A)^{-2}H_{Ferm}\Bigr)\nonumber\\
=N(\hbox{det}A)^{-1/2}\sqrt{\hbox{det}\widetilde{\Psi}}
\Bigl({1 \over G}\bigl(\Lambda+(\hbox{det}A)\hbox{tr}\widetilde{\Psi}^{-1}\bigr)+{{(\hbox{det}A)} \over {\hbox{det}\widetilde{\Psi}}}H_{Ferm}\Bigr).
\end{eqnarray}

\noindent
Next we will bring out a factor of $(\hbox{det}A)$ from the large brackets in (\ref{DYNAMICS2}), and taking $\widetilde{\psi}_{ae}$ without loss of generality to be diagonal, make the definitions

\begin{eqnarray}
\label{DYNAMICS21}
\widetilde{\Psi}_{33}=\lambda;~~\widetilde{\Psi}_{11}-\widetilde{\Psi}_{33}=\alpha;~~\widetilde{\Psi}_{22}-\widetilde{\Psi}_{33}=\beta.
\end{eqnarray}

\noindent
The variables $\alpha$, $\beta$ and $\lambda$ have been given the suggestive symbols of the eigenvalues of the gravitational states previously constructed.  But for the purposes of the present section these symbols are being used to denote classical momentum space variables.\footnote{The configuration space variables $(X,Y,T)$ retain the same status as in the previous section.}  Using (\ref{DYNAMICS21}), then (\ref{DYNAMICS2}) is given by

\begin{eqnarray}
\label{DYNAMICS3}
\boldsymbol{H}
={1 \over G}Na_0^{3/2}e^{T/2}\sqrt{\lambda(\lambda+\alpha)(\lambda+\beta)}\nonumber\\
\biggl[\Bigl(\Bigl({\Lambda \over {a_0^3}}\Bigr)e^{-T}
+\Bigl({1 \over \lambda}+{1 \over {\lambda+\alpha}}+{1 \over {\lambda+\beta}}\Bigr)\Bigr)\nonumber\\
+{1 \over {\lambda(\lambda+\alpha)(\lambda+\beta)}}\Bigl((3\lambda+\alpha+\beta)(GQ)+3i(GQ)^2\Bigr)\biggr].
\end{eqnarray}

\noindent
The Hamilton's equation of motion for the gravitational configuration variables are given by

\begin{eqnarray}
\label{DYNAMICS4}
\dot{X}=G{{\partial\boldsymbol{H}} \over {\partial\alpha}}
=Na_0^{3/2}e^{T/2}\sqrt{\lambda(\lambda+\alpha)(\lambda+\beta)}
\biggl[-\Bigl({1 \over {\lambda+\alpha}}\Bigr)^2-{1 \over {\lambda(\lambda+\alpha)^2(\lambda+\beta)}}\nonumber\\
\Bigl((3\lambda+\alpha+\beta)(GQ)+3i(GQ)^2\Bigr)
+{1 \over {\lambda(\lambda+\alpha)(\lambda+\beta)}}(GQ)\biggr],
\end{eqnarray}

\begin{eqnarray}
\label{DYNAMICS5}
\dot{Y}=G{{\partial\boldsymbol{H}} \over {\partial\beta}}
=Na_0^{3/2}e^{T/2}\sqrt{\lambda(\lambda+\alpha)(\lambda+\beta)}
\biggl[-\Bigl({1 \over {\lambda+\beta}}\Bigr)^2-{1 \over {\lambda(\lambda+\alpha)(\lambda+\beta)^2}}\nonumber\\
\Bigl((3\lambda+\alpha+\beta)(GQ)+3i(GQ)^2\Bigr)
+{1 \over {\lambda(\lambda+\alpha)(\lambda+\beta)}}(GQ)\biggr].
\end{eqnarray}

\noindent
and

\begin{eqnarray}
\label{DYNAMICS5}
\dot{T}=G{{\partial\boldsymbol{H}} \over {\partial\lambda}}
=Na_0^{3/2}e^{T/2}\sqrt{\lambda(\lambda+\alpha)(\lambda+\beta)}
\biggl[-\Bigl(\Bigl({1 \over \lambda}\Bigr)^2+({1 \over {\lambda+\alpha}}\Bigr)^2+\Bigl({1 \over {\lambda+\beta}}\Bigr)^2\Bigr)\nonumber\\
-\Bigl({1 \over {\lambda^2(\lambda+\alpha)(\lambda+\beta)}}+{1 \over {\lambda(\lambda+\alpha)^2(\lambda+\beta)}}+{1 \over {\lambda(\lambda+\alpha)(\lambda+\beta)^2}}\Bigr)\nonumber\\
\Bigl((3\lambda+\alpha+\beta)(GQ)+3i(GQ)^2\Bigr)
+{1 \over {\lambda(\lambda+\alpha)(\lambda+\beta)}}(GQ)\biggr].
\end{eqnarray}

\noindent
Transferring the exponential to the left hand side and using the identity

\begin{eqnarray}
\label{DYNAMICS51}
e^{-T/2}\dot{T}=-2{d \over {dt}}e^{-T/2},
\end{eqnarray}

\noindent
equation (\ref{DYNAMICS5}) can be written as

\begin{eqnarray}
\label{DYNAMICS51}
{d \over {dt}}e^{-T/2}=
-{1 \over 2}Na_0^{3/2}\sqrt{\lambda(\lambda+\alpha)(\lambda+\beta)}
\biggl[\Bigl(\Bigl({1 \over \lambda}\Bigr)^2+({1 \over {\lambda+\alpha}}\Bigr)^2+\Bigl({1 \over {\lambda+\beta}}\Bigr)^2\Bigr)\nonumber\\
\Bigl({1 \over {\lambda^2(\lambda+\alpha)(\lambda+\beta)}}+{1 \over {\lambda(\lambda+\alpha)^2(\lambda+\beta)}}+{1 \over {\lambda(\lambda+\alpha)(\lambda+\beta)^2}}\Bigr)\nonumber\\
-\Bigl((3\lambda+\alpha+\beta)(GQ)+3i(GQ)^2\Bigr)
-{1 \over {\lambda(\lambda+\alpha)(\lambda+\beta)}}(GQ)\biggr].
\end{eqnarray}

\noindent
The Hamiltonian constraint comes from the equation of motion for $N$

\begin{eqnarray}
\label{DYNAMICS6}
{{\partial\boldsymbol{H}} \over {\partial{N}}}
={1 \over G}a_0^{3/2}e^{T/2}\sqrt{\lambda(\lambda+\alpha)(\lambda+\beta)}\nonumber\\
\biggl[\Bigl(\Bigl({\Lambda \over {a_0^3}}\Bigr)e^{-T}
+\Bigl({1 \over \lambda}+{1 \over {\lambda+\alpha}}+{1 \over {\lambda+\beta}}\Bigr)\Bigr)\nonumber\\
+{1 \over {\lambda(\lambda+\alpha)(\lambda+\beta)}}\Bigl((3\lambda+\alpha+\beta)(GQ)+3i(GQ)^2\Bigr)\biggr]=0,
\end{eqnarray}

\noindent
which due to nondegeneracy of the pre-factors implies that

\begin{eqnarray}
\label{DYNAMICS7}
{1 \over {\lambda(\lambda+\alpha)(\lambda+\beta)}}
\Bigl((3\lambda+\alpha+\beta)(GQ)+3i(GQ)^2+{{G\pi^2} \over 2}\Bigr)\nonumber\\
=-\biggl[\Bigl({\Lambda \over {a_0^3}}\Bigr)e^{-T}+\Bigl({1 \over \lambda}+{1 \over {\lambda+\alpha}}+{1 \over {\lambda+\beta}}\Bigr)\biggr].
\end{eqnarray}

\noindent
Next we compute the Hamilton's equations of motion for the gravitational momenta.  These are given by

\begin{eqnarray}
\label{DYNAMICS8}
{1 \over G}\dot{\alpha}=-{{\partial\boldsymbol{H}} \over {\partial{X}}}=0;\nonumber\\
{1 \over G}\dot{\beta}=-{{\partial\boldsymbol{H}} \over {\partial{Y}}}=0;\nonumber\\
{1 \over G}\dot{\lambda}=-{{\partial\boldsymbol{H}} \over {\partial{T}}}
=-{1 \over G}Na_0^{3/2}e^{T/2}\sqrt{\lambda(\lambda+\alpha)(\lambda+\beta)}\nonumber\\
\biggl[-\Bigl({\Lambda \over {a_0^3}}\Bigr)e^{-T}+\Bigl({1 \over \lambda}+{1 \over {\lambda+\alpha}}+{1 \over {\lambda+\beta}}\Bigr)\nonumber\\
+{1 \over {\lambda(\lambda+\alpha)(\lambda+\beta)}}\Bigl((3\lambda+\alpha+\beta)(GQ)+3i(GQ)^2\Bigr)\biggr].
\end{eqnarray}

\noindent
From (\ref{DYNAMICS8}) one reads of that $\alpha$ and $\beta$ are arbitrary complex constants, independent of time.\footnote{This is consistent with the preservation in time of the gravitational coherent 
states since from (\ref{PRIORTO3}) and (\ref{PRIORTO31}), $\alpha$ and $\beta$ are directly related to the eigenvalue of the annihilation operators $a_1$ and $a_2$ on the states.}  Using the Hamiltonian constraint (\ref{DYNAMICS7}), the term in large brackets in the third equation of (\ref{DYNAMICS8}) is double the first term.  This reduces the equation of motion for $\lambda$ to

\begin{eqnarray}
\label{DYNAMICS9}
\dot{\lambda}=2\Lambda{a}_0^{-3/2}N\sqrt{\lambda(\lambda+\alpha)(\lambda+\beta)}e^{-T/2}.
\end{eqnarray}

\noindent
Hence in the instanton representation, all configuration space dependence within the Hamiltonian constraint is confined to $T$, which is a time variable.  In constrast to the case for $\alpha$ and $\beta$, it is inappropriate to regard $\lambda$ as a label corresponding to a coherent state, since it is not preserved in time by the equations of motion except when $\Lambda=0$.

\subsection{Equation of motion for the fermionic field}

\noindent
The equations of motion for the spin ${1 \over 2}$ fermion are given by

\begin{eqnarray}
\label{MASSLESS}
\dot{\psi}^A={{\partial\boldsymbol{H}} \over {\partial\pi_A}}=-{1 \over G}Na_0^{3/2}e^{T/2}{1 \over {\sqrt{\lambda(\lambda+\alpha)(\lambda+\beta)}}}G\Bigl((3\lambda+\alpha+\beta)+6i(GQ)\Bigr)\psi^A;\nonumber\\
\dot{\pi}_A=-{{\partial\boldsymbol{H}} \over {\partial\psi^A}}={1 \over G}Na_0^{3/2}e^{T/2}{1 \over {\sqrt{\lambda(\lambda+\alpha)(\lambda+\beta)}}}G\Bigl((3\lambda+\alpha+\beta)+6i(GQ)\Bigr)\pi_A
\end{eqnarray}

\noindent
Note that

\begin{eqnarray}
\label{MASSLESS1}
\dot{Q}=\pi_A\dot{\psi}^A+\dot{\pi}_A\psi^A.
\end{eqnarray}

\noindent
Hence the following equation is implied by the equations of motion (\ref{MASSLESS})

\begin{eqnarray}
\label{MASSLESS2}
\dot{Q}={1 \over G}Na_0^{3/2}e^{T/2}{1 \over {\sqrt{\lambda(\lambda+\alpha)(\lambda+\beta)}}}\nonumber\\
G\Bigl((3\lambda+\alpha+\beta)+6i(GQ)\Bigr)\Bigl((3\lambda+\alpha+\beta)(GQ)+3i(GQ)^2\Bigr),
\end{eqnarray}

\noindent
which features the time evolution of $Q$ in the gravitational background labelled by $\alpha$ and $\beta$.  Equation (\ref{DYNAMICS4}), (\ref{DYNAMICS5}), (\ref{DYNAMICS6}), (\ref{DYNAMICS9}) and (\ref{MASSLESS2}) form a closed system and will form the subject of future study.  The main physical content which we wanted to get out of the Hamilton's equations is a consistency check on the quantum dynamics.  The check is that the 
gravitational variables $(\alpha,\beta)$ remain stationary in spite of the presence of the fermion, and appropriately label the quantum state.

\section{Conclusion}

In this paper we have demonstrated the existence of a well-defined Hilbert space for gravity coupled to fermions with a good semiclassical limit.  This paper may be taken as a first step in the quantization of the Standard Model, taking gravity into account.  The results of matter coupling appears to not be inconsistent with the instanton representation of gravity, at least in minisuperspace.  A future step in this direction will be the inclusion of additional fields with the fermions, starting with the scalar field and moving on to include the Yang--Mills field.  The ability to construct the analogous Hilbert space for all three fields combined should provide a starting point for the investigation of the Standard Model.

\end{document}